\documentclass[twocolumn]{osa-article}

\usepackage{cite}
\journal{oe}
\articletype{Research Article}

\usepackage{lineno}
\usepackage{gensymb}

\begin{document}

\title{High finesse bow-tie cavity for strong atom-photon coupling in Rydberg arrays}

\author{Yu-Ting Chen\authormark{*1,2}, Michal Szurek\authormark{1}, Beili Hu\authormark{1}, Julius de Hond\authormark{1}, Boris Braverman\authormark{3}, and Vladan Vuletic\authormark{1}}

\address{\authormark{1}Department of Physics and Research Laboratory of Electronics, Massachusetts Institute of Technology, 77 Massachusetts Avenue, Cambridge, MA 02139, USA\\
\authormark{2}Department of Physics, Harvard University, 17 Oxford Street, Cambridge, MA 02138, USA\\\authormark{3}Department of Physics, University of Ottawa, 25 Templeton Street, Ottawa, ON, K1N 6N5, Canada}

\email{\authormark{*}yutingchen@g.harvard.edu}

\begin{abstract*}

We report a high-finesse bow-tie cavity designed for atomic physics experiments with Rydberg atom arrays. The cavity has a finesse of 51,000 and a waist of 7.1~$\mu$m at the cesium D2 line (852 nm). With these parameters, the cavity is expected to induce strong coupling between a single atom and a single photon, corresponding to a cooperativity per traveling mode of 35 at the cavity waist. To trap and image atoms, the cavity setup utilizes two in-vacuum aspheric lenses with a numerical aperture ($NA$) of 0.35 and is capable of housing $NA = 0.5$ microscope objectives. In addition, the large atom-mirror distance ($\gtrsim1.5$~cm) provides good optical access and minimizes stray electric fields at the position of the atoms. This cavity setup can operate in tandem with a Rydberg array platform, creating a fully connected system for quantum simulation and computation.
\end{abstract*}

\section{Introduction}
Bow-tie cavities are widely used in laser technology. Such cavities support two counter-propagating unidirectional modes, each with one small and one large mode waist, and are widely used for pumped lasers and as frequency-doubling cavities. Besides applications in laser physics, bow-tie cavities have recently found use in atomic-physics experiments, where the traveling modes facilitate uniform light-matter interactions \cite{jia_strongly_2018}. In these experiments, the cavity and the atoms couple through the coherent exchange of photons. Two conditions must be met to create strong light-atom coupling, i.e., the coherent exchange of an excitation between a single atom and the cavity mode \cite{tanji-suzuki_interaction_2011}. First, the light needs to travel between mirrors numerous times to increase the chance of being absorbed or emitted by the atom. Second, the mode waist needs to be small to maximize the light intensity at the atoms. The former can be quantified by the cavity finesse $\mathcal{F}$; the latter can be quantified by the cavity small mode waist $w$.

The cavity cooperativity $\eta$ is a dimensionless parameter that can be used to quantify the light-atom interactions. It describes the ratio of atom emission into the cavity relative to the emission into free space \cite{tanji-suzuki_interaction_2011}. The cooperativity is proportional to the finesse $\mathcal{F}$ of the cavity and inversely proportional to the square of the small mode waist $w$. For a standing-wave mode, the peak cooperativity is $\eta = 24\mathcal{F}/(\pi k^{2}w^{2})$, where $k=2\pi/\lambda$ is the wave vector of the light \cite{tanji-suzuki_interaction_2011}. Larger cooperativity corresponds to stronger light-atom coupling, and $\eta \geq 1$ marks the transition to the strong-coupling regime. Strong light-atom coupling opens up new possibilities for research and technology in quantum optics \cite{obrien_photonic_2009,welte_cnot_2018} and entanglement generation \cite{schleier-smith_squeezing_2010,McConnell_2015,davis_heisenberg_2016}. Furthermore, introducing Rydberg atom arrays \cite{barredo_atom-by-atom_2016,browaeys_many-body_2020,ebadi_quantum_2021} into the cavity enables the study of long-range spin physics and potentially fully connected quantum computation \cite{periwal_programmable_2021}.  

Loading Rydberg atoms into a cavity benefits from large optical access. However, it is demanding to build a cavity that simultaneously features high cooperativity, large optical access, and high stability. Among the existing cavity quantum electrodynamics (cavity QED) setups that strongly couple light to individual atoms, most are designed with a mirror in close proximity ($\lesssim500$ $\mu$m) to the atoms in order to minimize the cavity waist at the atoms \cite{thompson_observation_1992, hunger_fiber_2010,kawasaki_geometrically_2019,liu_cavity_2022}. However, implementing Rydberg arrays into these setups poses significant challenges due to the limited optical access and the sensitivity of Rydberg atoms to stray electric fields from nearby mirror surfaces \cite{abel_rydberg_2011,sedlacek_quartz_2016,davtyan_chip_2018}. One exception to this general trend is a near-concentric cavity, where the separation of two spherical mirrors is nearly equal to twice their radius of curvature \cite{deist_superresolution_2022}. A near-concentric cavity provides large optical access, but it operates close to an unstable regime and is far more sensitive to angular misalignment. Therefore, it is of interest to explore different cavity geometries to overcome the challenge of constructing a stable cavity with high cooperativity and large optical access.

In this article, we present a bow-tie cavity that offers high cooperativity, large mirror spacing, and good mechanical stability. The cavity is composed of two concave mirrors and two convex mirrors, generating a TEM$_{00}$ mode waist $w=7.1$ $\mu$m  between the two concave mirrors. With a finesse $\mathcal{F}=5.1 \times 10^4$, the cavity is expected to have peak cooperativity of $\eta_1=35$ for a traveling-wave mode ($\eta_2=140$ peak cooperativity for a standing-wave mode with the factor of 4 coming from the constructive interference). The cavity is mechanically stable, with a $3.1$-cm spacing between the concave mirrors providing good optical access and reducing the stray electric field at the position of the atoms near the cavity waist. For trapping and imaging atoms, two aspheric lenses ($NA = 0.35$) are aligned to image the region near the cavity waist. The whole cavity setup is operated under vacuum ($3.6\times 10^{-9}$ Torr).

\begin{figure}[!ht]
\graphicspath{{Figures/}}
\centering\includegraphics[width=8.4cm]{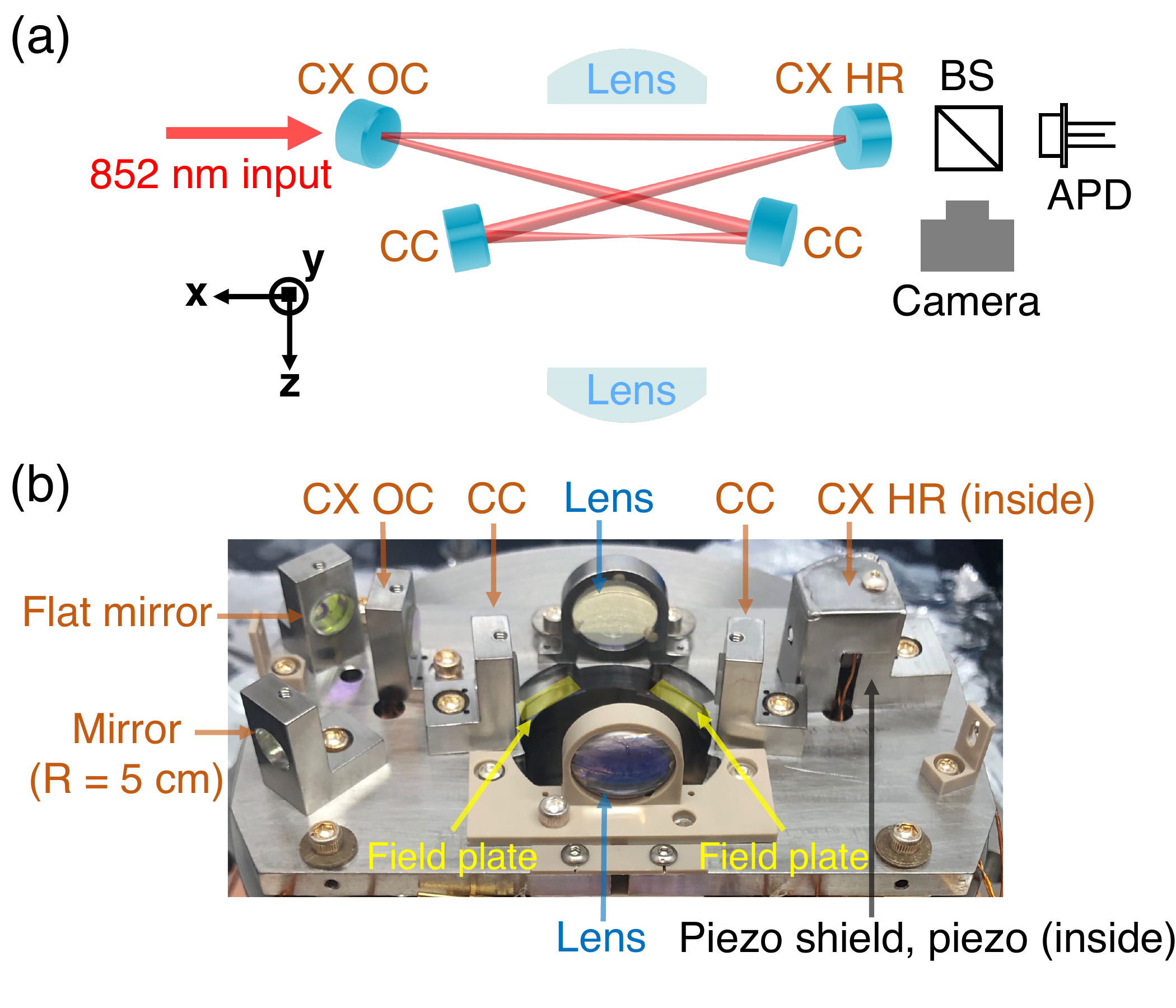}
\caption{Experimental setup. (a) Overview of the cavity, including the two concave (CC) mirrors, a convex output coupler (CX OC) mirror, a convex high reflector (CX HR) mirror, and two aspheric lenses. The atoms are placed between the two concave mirrors where the mode is the smallest. At the output of the cavity, a beamsplitter (BS) divides the signal to an avalanche photodiode (APD) and a camera. The APD records the signal for finesse measurement, and the camera is used for mode size measurements and initial alignment to the TEM$_{00}$ mode. (b) Photograph of the bow-tie cavity setup, showing the mirrors and aspheric lenses inside their holders, the metallic shield covering the piezoelectric mirror transducer, and two of the eight electric-field plates. All components are mounted on a stainless-steel base plate with silver-plated screws.}
\label{fig:setup}
\end{figure}

\section{Cavity design}

\begin{figure}[!ht]
\graphicspath{{Figures/}}
\centering\includegraphics[width=13cm]{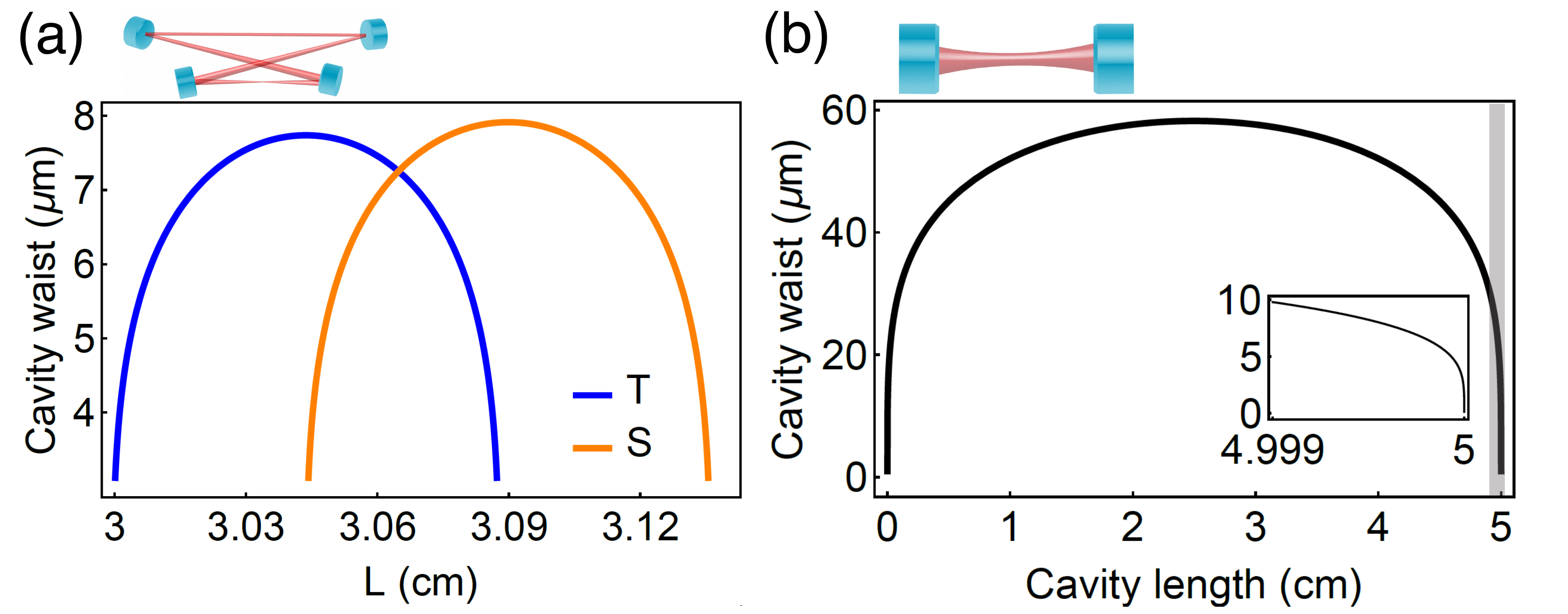}
\caption{Cavity mode waist versus cavity length. (a) Cavity stability plot of the bow-tie cavity. The cavity is stable when the tangential (blue) and sagittal (orange) waists co-exist. L: distance between the two concave mirrors. (b) Cavity stability plot of a two-mirror cavity with $2.5$ cm mirror curvature. The gray shaded area marks the near-concentric region. Inset: Zoom-in to the near-concentric region. In contrast to the bow-tie cavity shown on the left, this near-concentric cavity needs to be stabilized within a 10 $\mu$m range of cavity lengths to obtain a waist of less than 10 $\mu$m, which is challenging to adjust and maintain during the baking of the vacuum chamber.}
\label{fig:calculatedWaist}
\end{figure}

We performed ABCD matrix calculations \cite{siegman,nagourney_quantum_2014} to design a cavity meeting all three requirements: small mode waist, large mirror separations, and high alignment stability. The cavity is composed of two concave mirrors with $R_1=2.46$~cm radius of curvature and two convex mirrors with $R_2=-3.87$~cm radius of curvature. The two concave mirrors are separated by $3.1$ cm; the two convex mirrors are separated by $6.1$ cm. The cavity mode has a $\theta=6.42\degree$ incidence angle on all mirrors, and all mirrors are in the same plane. The two convex mirrors serve to expand the laser beam on the concave mirrors, minimizing the size of the small mode waist \cite{jia_strongly_2018}. The cavity stability plot is shown in Fig.~\ref{fig:calculatedWaist}(a). The two curves correspond to the tangential and sagittal waists, which are separated due to astigmatism caused by the incidence angle $\theta$. The cavity is designed with its length equal to the crossing point of the two curves for the tangential and sagittal waists in Fig.~\ref{fig:calculatedWaist}(a), maximizing cavity stability under changes in the cavity length. A 7-$\mu$m waist is the smallest achievable as set by the length of the stability region for this mirror separation. The mode waist could be reduced by either further separating the convex mirrors (while fixing the mirror curvatures) or by reducing the curvature of the concave mirrors while moving these mirrors closer together. The downside of the former approach is that it would narrow the stability window, and the downside of the latter is that it decreases the distance between the Rydberg atoms and the nearest dielectric surface. We emphasize that this bow-tie cavity has higher stability compared to a two-mirror cavity with similar waist size and mirror spacing. The two-mirror cavity would need to be operated in the near-concentric regime, in which the mode is sensitive to the mirror spacing and tilts. Figure~\ref{fig:calculatedWaist}(b) shows the stability curve for a near-concentric cavity.

All mirrors are coated for applications at the cesium D1 (895~nm) and D2 (852~nm) transitions. A low-transmission ($2.5$ ppm) coating is used on both concave mirrors and one of the convex mirrors (convex high reflector, CX HR) to maximize the cavity finesse. An intermediate-transmission ($54$ ppm) coating is used on the other convex mirror (convex output coupler, CX OC) to achieve a cavity out-coupling efficiency of 44\% given a finesse $5.1\times10^{4}$. In addition to the D1 and D2 lines, all four mirrors are coated at the wavelength corresponding to the transition from the cesium $P_{3/2}$ state to a typical Rydberg level (508 nm), with 1000 ppm mirror transmission.

\section{Cavity assembly}\label{assembly}
The cavity structure supports precise mirror alignment and is vacuum compatible. The structure consists of four mirror holders and a base plate, all made of non-magnetic 316 stainless steel with high machining precision (tolerance $5$ - $13$ $\mu$m). The mirror holders are mounted to the base plate with two dowel pins to guide the alignment. Inside the mirror holders, mirrors are mounted to a precise three-point-contact structure, with two points on the holder and one point defined by a silver-tip set screw. The cavity assembly consisted of initial and final alignment procedures. In the initial alignment, four mirrors were aligned with the aid of kinematic mirror mounts (Thorlabs MK05). In the final alignment, the kinematic mounts were replaced by stainless mounts one by one for vacuum compatibility and mechanical stability. The assembly was performed by tuning the angles of the convex mirrors and the spatial mode of the input laser. The two concave mirrors remained fixed throughout the alignment, with their positions defined by dowel pins on the base plate. To form a cavity mode, the angles of the two convex mirrors were adjusted iteratively until the laser beam overlapped with itself after a cavity round trip.

After the final alignment, all mirrors were glued to stainless-steel holders to provide vacuum compatibility and cavity stability. The two Thorlabs MK05 holders were replaced by stainless mirror holders one at a time, with one convex mirror (CX HR) glued onto a piezo ring actuator (Noliac NAC2123) for actively stabilizing the cavity length. During this process, a camera constantly monitored the cavity mode at the cavity output. In addition, two mirrors (one flat mirror and one concave mirror with a 5 cm radius of curvature) were installed to guide and collimate the output beam from the cavity. After the alignment was finished, all mirrors were glued to the stainless steel holders with vacuum-compatible epoxy (Epoxy Technology H77), and then all silver-tip set screws were removed for vacuum compatibility. This procedure completed the assembly of the bow-tie cavity. A photograph of the cavity setup is presented in Fig. \ref{fig:setup}(b).

\begin{figure}[!ht]
\graphicspath{{Figures/}}
\centering\includegraphics[width = 8.4cm]{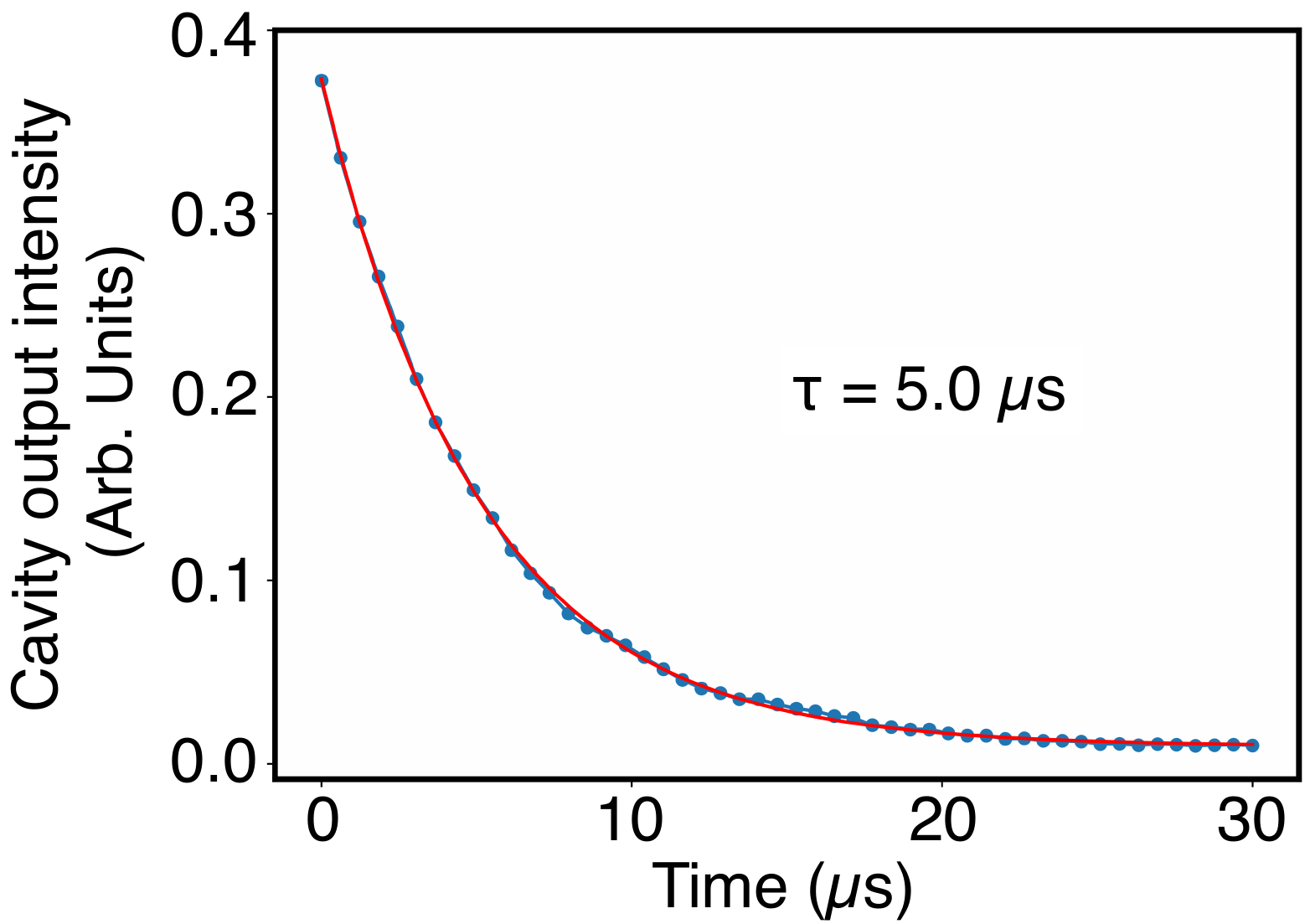}
\caption{Cavity ringdown measurement to determine the cavity finesse. During the ringdown measurement, the input beam was switched off, and the decay of the cavity light was measured with an avalanche photodiode. The decay curve was fitted to an exponential function. In this measurement, the 5.0 $\mu$s photon decay time, together with the 18.6 cm cavity length, corresponds to a finesse of $\mathcal{F} = 5.1 \times 10^4$. Each data point in this plot is an average over $0.6$ $\mu$s period to increase the signal-to-noise ratio.}
\label{fig:ringdown}
\end{figure}

Cavity finesse and waist size decide the cavity cooperativity for a particular atomic transition. We measured the cavity finesse with the ringdown method using an avalanche photodiode (Thorlabs APD$120$A)\cite{nagourney_quantum_2014}, as depicted in Fig. \ref{fig:setup}(a). Figure \ref{fig:ringdown} shows a ringdown measurement result. The final finesse is $\mathcal{F}=5.1\times 10^4$ after baking the vacuum chamber. It is worthwhile mentioning that the maximum measured finesse was even higher ($6.6\times 10^4$) when the two convex mirrors were mounted on kinematic mounts due to the additional degrees of freedom that facilitated alignment.

\begin{figure}[!ht]
\graphicspath{{Figures/}}
\centering\includegraphics[width=8.4cm]{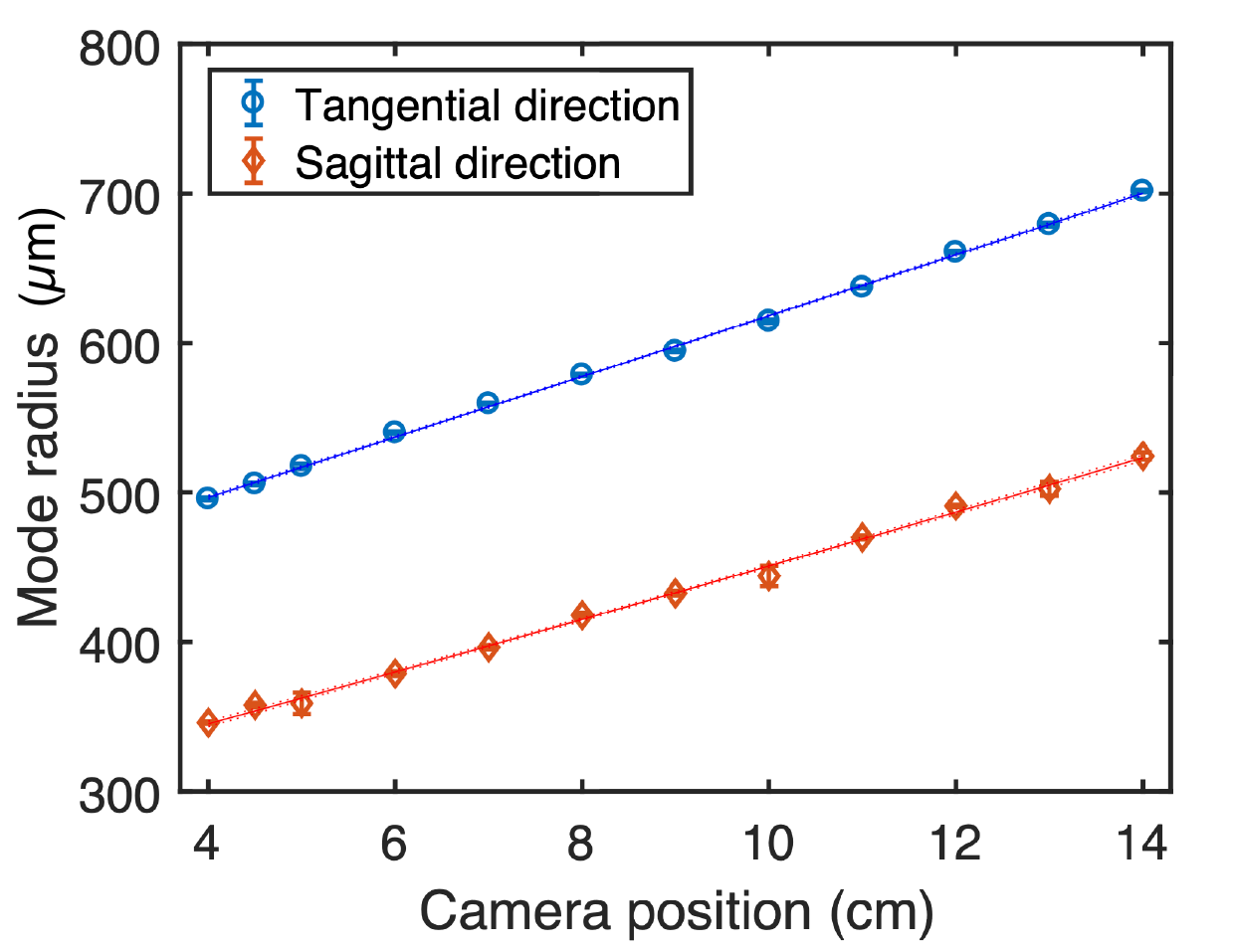}
\caption{Characterization of the cavity mode. The size of the cavity waist is derived from the divergence of the mode outside the cavity, yielding tangential and sagittal waists of $w_t=7.1$ $\mu$m and $w_s=7.2$ $\mu$m, respectively.}
\label{fig:waistMeasurement}
\end{figure}

To determine the sizes of the tangential ($w_t$) and sagittal ($w_s$) cavity mode waists, we measured the expansion of the mode outside the cavity, as illustrated in Fig.~\ref{fig:waistMeasurement}. We then used ABCD matrices to propagate the laser beam back into the cavity and calculate the waist size. The calculation results in $w_t=7.1$ $\mu$m and $w_s=7.2$ $\mu$m, which agrees with the design (Fig.~\ref{fig:calculatedWaist}). 

\begin{figure}[!ht]
\graphicspath{{Figures/}}
\centering\includegraphics[width=7cm]{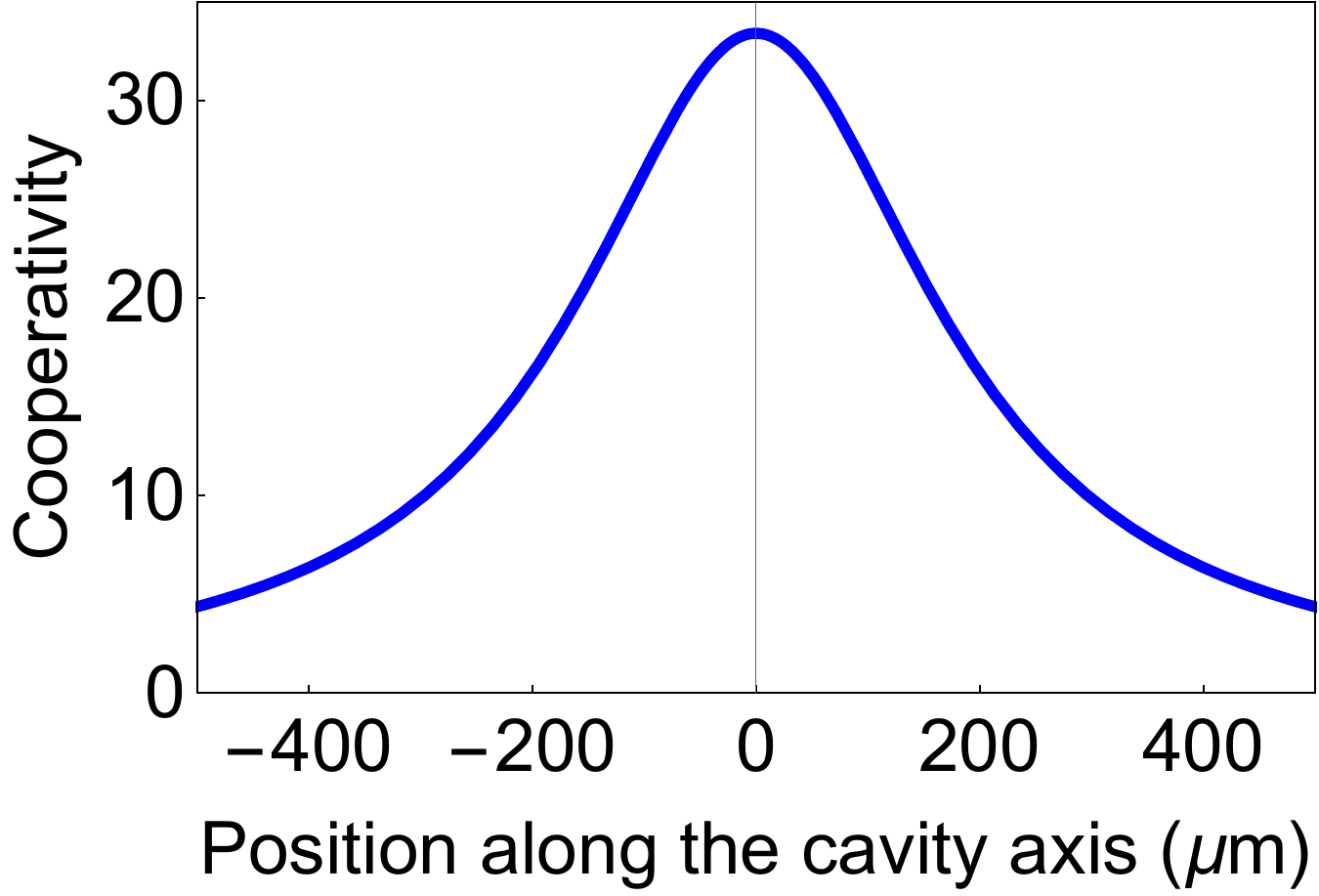}
\caption{Averaged single-atom cooperativity along the cavity axis per running-wave mode. The averaged single-atom cooperativity remains higher than 15 within a range of $\pm 200$ $\mu$m, showing that the cavity can mediate strong light-atom couplings within the typical size of an atom array.}
\label{fig:cooperativity}
\end{figure}

Given the finesse and the waist, we can derive the maximum cooperativity per traveling-wave mode as $\eta_1= 6\mathcal{F}/(\pi k^2 w_t w_s)=35$, where $k=2\pi/ \lambda$ with $\lambda = 852$~nm is the wavelength for the cesium D2 line. This high cooperativity enables strong light-atom interactions. Figure~\ref{fig:cooperativity} shows the average cooperativity along the cavity axis. The averaged cooperativity remains above 15 within a $\pm 200$ $\mu$m range, showing that this cavity is suitable for placing atom arrays, e.g., cesium atom arrays with $4$ $\mu$m atomic spacing, along the cavity axis to study light-mediated interactions. 

The single-atom cooperativity $\eta_1$ corresponds to a coupling of the atom to a single traveling mode, while the maximum for coupling to both modes is $\eta_2=4\eta_1=140$. Experimentally, the traveling-wave cooperativity is realized when the atom is coupled to a laser traveling along the cavity in one direction. On the other hand, the two-mode cooperativity is realized when two laser beams are sent to the cavity simultaneously or when the atom is excited from the side and emits a photon simultaneously into both directions of the bow-tie cavity.

Bow-tie cavities intrinsically have birefringence due to the non-zero angle between the cavity mode and mirror surfaces. The two orthogonal polarizations experience different boundary conditions on the mirror surfaces, potentially affecting the finesse and the resonance frequency of the two polarization modes. In this cavity, the s and p polarizations have the same finesse due to the small $6.42\degree$ laser angle of incidence. However, the resonance frequencies of the two polarizations differ by more than one cavity linewidth, indicating that this cavity can only support linearly polarized light. Therefore, when the cavity is coupled to atoms, the cooperativity needs to be modified based on the polarization and the oscillator strength of the atomic transitions coupled to the cavity. For a given transition, the effective cooperativity would be scaled by the squared Clebsch-Gordan coefficient of the atomic transition and an additional factor of $1/2$ if using a $\sigma+$ or $\sigma-$ transition. For example, the scaling factors for a few of the Cs D2 transitions are: $0.56$ for $|3,3\rangle\rightarrow|3',3'\rangle$; $0.21$ for $|3,3\rangle\rightarrow|4',4'\rangle$; $0.47$ for $|4,4\rangle\rightarrow|4',4'\rangle$; $0.5$ for $|4,4\rangle\rightarrow|5',5'\rangle$.

We implement several additional features to maintain the mechanical and electrical stability of the cavity. For mechanical stability, two heaters (Allectra 343-HEATER-2X10-V2) and two thermistors (TDK B57550G1104F000) are glued onto the back of the base plate to stabilize the cavity length and to fine-tune the cavity alignment by setting one side of the cavity base plate at a higher temperature. During heating, the cavity temperature is kept under $30^{\circ}$C to minimize vacuum outgassing. To stabilize the electric fields experienced by the atoms, eight arc-shaped field-plates were mounted around the cavity waist, and the piezoelectric transducer for the cavity mirror is shielded by a metallic cover \cite{low_experimental_2012,YT_2022}. The cavity structure is also electrically grounded.

\section{Lens alignment}
In addition to the bow-tie cavity, this setup contains two in-vacuum aspheric lenses (Edmund Optics 49-111) to trap and image atoms at the cavity waist. In order to align the lenses to the cavity mode waist, we created a 200-nm aperture point source at the cavity waist out of a scanning near-field optical microscopy (SNOM) tip \cite{ferri_mapping_2020}. The alignment procedure includes: (1) selecting a few positions along the cavity axis ($x$-axis) which are close to the cavity center; (2) at each selected $x$ position, moving the SNOM tip in the $yz$-plane to find the mode center and to estimate the mode radius; (3) finding the $x$ coordinate of the cavity waist $(x_{c})$, which is the $x$ position with the smallest mode radius; (4) moving the SNOM tip to $x_{c}$ and its corresponding mode center acquired in step 2. After these four steps, the SNOM tip is located at the cavity mode waist $(x_{c}, y_{c}, z_{c})$. 

\begin{figure}[!ht]
\graphicspath{{Figures/}}
\centering\includegraphics[width=13cm]{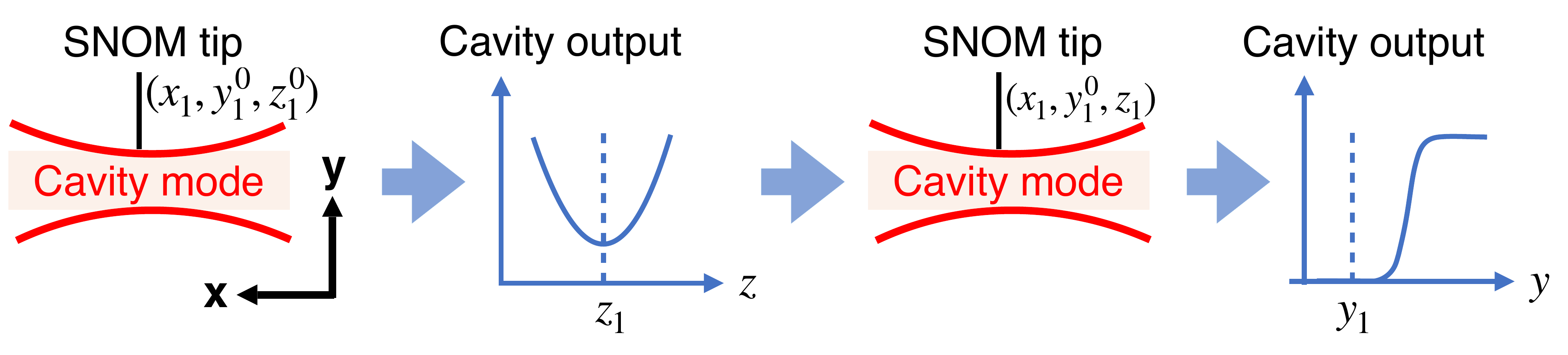}
\caption{Diagram summarizing the technique for finding the cavity waist position using a SNOM tip.}
\label{fig:SNOMScanzy}
\end{figure}

Figure~\ref{fig:SNOMScanzy} shows the procedure for positioning the SNOM tip to the cavity mode waist. This process was done using a 3D translation stage while monitoring the cavity output intensity. The SNOM tip was initially placed at position $(x_{1}, y^{0}_{1}, z^{0}_{1})$, with $x_{1}$ close to the cavity center. To find the mode center at $x_{1}$, the SNOM tip was first scanned in the $z$ direction across the cavity mode while the transmission was monitored. When the SNOM tip was moved toward the mode center, the intensity decreased. The $z$ coordinate of the mode center $(z_{1})$ was thus determined from the minimum of the transmission. Subsequently, the same procedure was repeated for the $y$ axis to find the mode center $(y_{1})$ along $y$.

\begin{figure}[!ht]
\graphicspath{{Figures/}}
\centering\includegraphics[width=9cm]{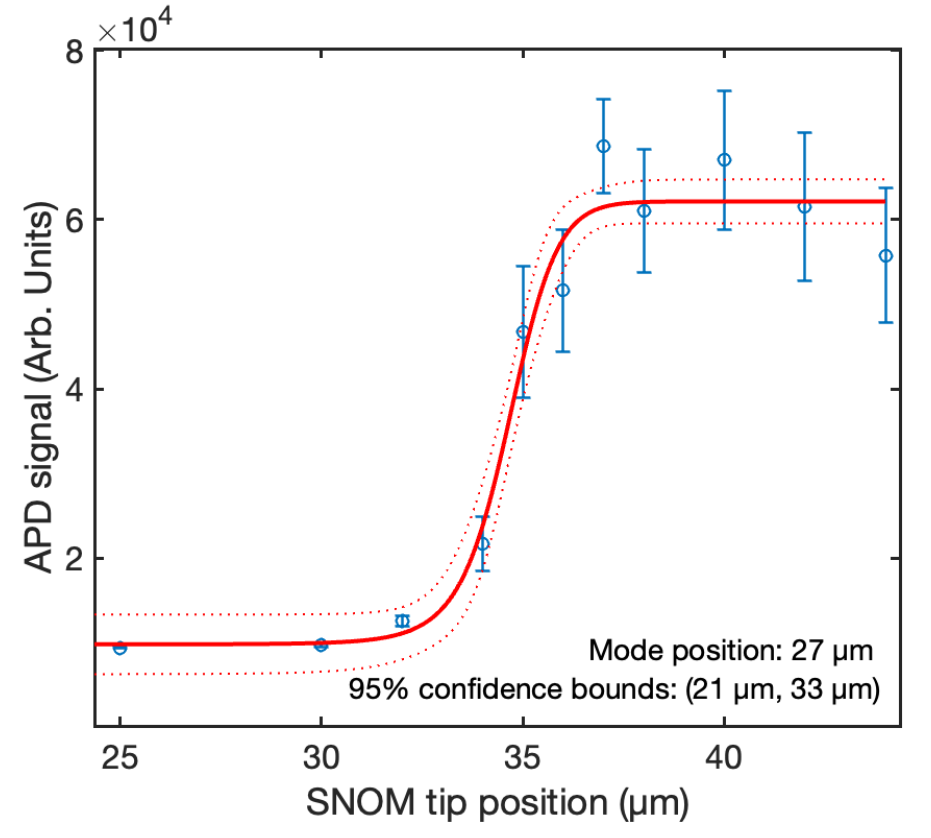}
\caption{Transmission through the cavity as the SNOM tip is scanned across the cavity mode. As the SNOM tip is moved along $y$, it induces cavity loss as light is scattered out of the cavity. The waist size can be estimated by fitting cavity output intensity with Eq. (\ref{lossesToOutput}). The dashed lines show the 68\% prediction interval, i.e., a new data point has a 68\% chance to lie inside the region between the two dashed lines.}
\label{fig:SNOMScan}
\end{figure}

For a cylindrical tip at position $y'$, the losses $L$ can be modeled as the percentage of the Gaussian beam area blocked by the SNOM tip:
\begin{equation}\label{intensity}
\mathbf{L}_{y'-y_{1}} =  \frac{\int_{-d/2}^{d/2}  \int_{y'-y_{1}}^{\infty} e^{-2(x^{2} + y^{2})/w_{1}^{2}} dx\,dy} {\int_{-\infty}^{\infty}  \int_{-\infty}^{\infty} e^{-2(x^{2} + y^{2})/w_{1}^{2}} dx\,dy}= \frac{1}{2} \mathrm{erf}\left(\frac{d}{\sqrt{2}w_{1}}\right)\mathrm{erfc}\left(\sqrt{2}\frac{y'-y_{1}}{w_{1}}\right), 
\end{equation}
where $d$ is the tip diameter, $y_{1}$ is the mode position, and $w_{1}$ is the mode radius. These extra losses reduce the effective cavity finesse. Therefore, the cavity output intensity dropped as the SNOM tip entered the mode, as illustrated in Fig. \ref{fig:SNOMScan}. The cavity output intensity $\mathbf{O}$ can be modeled as
\begin{equation}\label{lossesToOutput}
\mathbf{O}_{y'-y_{1}}=I_{0}+\frac{I}{(T_{c}+L_{c})+\mathbf{L}_{y'-y_{1}}},
\end{equation}
where $I_{0}$ is the background intensity, $I$ is the intensity in the mode, $T_{c}$ is the total mirror transmission, and $L_{c}$ is the total cavity losses. $T_{c}+L_{c}$ can be obtained using the formula $T_{c}+L_{c} = 2\pi/\mathcal{F}$, where $\mathcal{F}$ is the cavity finesse.

By fitting the cavity output with Eq. (\ref{lossesToOutput}), the $y$ position of the mode center $y_{1}$ can be found, and the mode radius $w_{1}$ can be estimated. We repeated this procedure for each selected $x$ position, then selected the $x$ position with the smallest mode radius, and aligned the SNOM tip to the cavity waist $(x_{c}, y_{c}, z_{c})$. 

Once the SNOM tip was aligned to the cavity waist, a laser beam traveling through the fiber and emerging from the SNOM tip was used as a point source for lens alignment. We aligned the lenses by optimizing the point spread functions imaged through the lenses, achieving $2.5$ $\mu$m resolution measured with $852$ nm. (The deviation from the diffraction limit likely comes from the manufacturing imperfection of the off-the-shelf lenses.) Additionally, to ensure the focal planes of the lenses overlapped with the cavity axis, the point spread functions were also optimized as the SNOM tip moved by $\pm 100$ $\mu$m along the cavity axis. This additional alignment allows the lenses to trap and image atom arrays along the cavity mode. After the lens alignment was finished, both lenses were glued onto the lens mounts with vacuum-compatible epoxy.

\section{Conclusion}
In conclusion, we have developed a bow-tie cavity with a small cavity waist, a large mirror separation, and high mechanical stability for atomic physics experiments. The cavity is expected to have an average cooperativity per traveling mode of $35$, corresponding to strong coupling between a single atom and a single photon in the cavity mode. The bow-tie geometry offers a homogeneous light-atom coupling and thus provides a platform to explore physics different from the standard standing-wave cavities \cite{ostermann2022superglass}. For trapping and imaging atoms, two high $NA$ aspheric lenses are aligned to the cavity waist, and the cavity is also compatible with high $NA$ microscope objectives, enabling an array of more than 50 individually controlled tweezer traps inside the cavity mode and potentially having two rows of traps side by side. Once the strong coupling is confirmed with atoms, this new cavity setup will open up research directions in entangled-state generation \cite{chen_carving_2015,schupp_2021}, quantum optics \cite{PhysRevX.12.021034}, quantum simulation of long-range spin physics \cite{lucas_NP_2014,smith_many-body_2016,belyansky_minimal_2020,PhysRevA.105.L041302,roux_strongly_2020}, and quantum computation with Rydberg atom arrays \cite{ramette_teleportation_2021}, pushing the boundary of quantum sensing and advancing our understanding of fundamental physics.

\begin{backmatter}
\bmsection{Funding}
National Science Foundation (Frontier Center program, Award \# PHY-1734011 and Quantum Leap Challenge Institutes Program, Award \# PHY-2016244).

\bmsection{Acknowledgments}
Yu-Ting Chen would like to acknowledge her mentors for their kind support as well as the many wonderful researchers with whom she had fruitful discussions, including but not limited to: Ningyuan Jia, Nathan Schine, Jonathan Simon, Woo Chang Chung, Jinggang Xiang, Alban Urvoy, Zachary Vendeiro, Jessica Y.-C. Yeh, Pablo Solano, Wenchao Xu, Aditya Venkatramani, Sergio Cantu, Julian Leonard, Annie Park, Andre Heinz, and more. This work was made possible by the excellent machinists Andrew Gallant, Michael Abruzzese, and Mark Belanger.

\bmsection{Disclosures}
The authors declare no conflicts of interest.

\bmsection{Data availability} Data underlying the results presented in this paper are not publicly available at this time but may be obtained from the authors upon reasonable request.

\end{backmatter}

\bibliography{A_high_finesse_bow_tie_cavity_for_strong_atom_photon_coupling_in_Rydberg_arrays}

\end{document}